\documentclass[%
  reprint,
unsortedaddress,
aps,
]{revtex4-2}

\usepackage{graphicx}
\usepackage{dcolumn}
\usepackage{bm}
\usepackage{nomencl}
\usepackage{amssymb}
\usepackage{etoolbox}
\usepackage{epsfig}
\usepackage{epstopdf}

\usepackage{hyperref}

\usepackage{xcolor}
\usepackage[version=4]{mhchem}

\begin{document}

\preprint{}

\title{Energy-based interpretation of the dispersion coefficient of the constant phase element}

\author{Anis Allagui$^*$}
\email{aallagui@sharjah.ac.ae}

\affiliation{Dept. of Sustainable and Renewable Energy Engineering, University of Sharjah, Sharjah 27272, United Arab Emirates}

\altaffiliation[Also at ]{Center for Advanced Materials Research, Research Institute of Sciences and Engineering, University of Sharjah, Sharjah 27272,  United Arab Emirates}

\affiliation{Dept. of Electrical and Computer Engineering, Florida International University, Miami, FL33174, United States}

\author{Enrique H. Balaguera}
\affiliation{
Escuela Superior de Ciencias Experimentales y Tecnología, Universidad Rey Juan
Carlos, C/ Tulip\'{a}n, s/n, 28933 M\'{o}stoles, Madrid, Spain
}

\author{Ahmed Elwakil}
\affiliation{Dept. of Electrical Engineering, University of Sharjah, Sharjah 27272, United Arab Emirates}

\affiliation{Nanoelectronics Integrated Systems Center, Nile University, Cairo 12588, Egypt}

\affiliation{Dept. of Electrical and Software Engineering, University of Calgary, Calgary, Alberta T2N 1N4, Canada}

\begin{abstract}

The dispersion coefficient of the constant phase element (CPE) is typically treated as an empirical fitting parameter in the analysis of impedance spectroscopy data, with no clear physical meaning. Here we seek to establish a   energy-based interpretation for this   coefficient by linking it to the ratio of the dissipated or stored energy in the CPE relative to that supplied by the input source. Using the $RC$ network equivalency of a CPE, we  decompose the total input energy  into a contribution stored in the capacitive modes and another dissipated in the resistive modes. Analytical expressions are   derived for three test examples: (i) a constant voltage, (ii) a voltage ramp, and (iii) a quadratic input of the form $v(t)=\lambda t^2$.   In all cases we found that the ratios of any two of these energy quantities reduce to  pure functions of the dispersion coefficient of the CPE, independent of excitation amplitude or material parameters. This result provides a new perspective  of the  CPE's dispersion coefficient  from a thermodynamic/energetic basis, with direct implications for supercapacitor characterization, battery modeling, as well as for the analysis of other electrochemical systems and devices exhibiting the CPE behavior.

\end{abstract}

\keywords{Constant phase element, Impedance, Mittag-Leffler function, Supercapacitors}
\maketitle

\section{Introduction}
\label{sec:introduction}

The constant phase element (CPE) is undeniably a standard tool regularly used for the analysis and interpretation of impedance spectroscopy data of practically most real electrochemical interfaces\;\cite{gateman2022use}. Numerous origins have been proposed for the non-ideal  behavior of the CPE, including (i)    roughness, porosity and fractal morphology of the electrode surface, (ii) non-uniform  distribution of  double-layer capacitance and charge transfer resistance at the electrode/electrolyte interface, as well as (iii) chemical heterogeneities resulting from adsorption or solution impurities\;\;\cite{gateman2022use, lasia2022origin, cordoba2015relationship}.  
However, despite the considerable  efforts that have been invested in  the experimental and theoretical studies of    systems exhibiting the  CPE behavior, a clear physical meaning of the element and its parameters is still clouded and ill-defined\;\cite{cordoba2015relationship, lasia2022origin}.

We recall that the CPE is defined from its impedance  function   
$
\tilde{z}(s)
  = {1}/({c_{\alpha} s^{\alpha}})
$ 
 where 
 $\alpha\;(0<\alpha<1)$ is known at the dispersion coefficient of the CPE, $c_{\alpha}$  as a pseudocapacitance (positive) constant  in units of F\,s$^{\alpha-1}$, and 
  $s^{\alpha}$ is equal to $\omega^{\alpha} (\cos(\alpha\pi/2) + j \sin(\alpha\pi/2))$.  The impedance, having both its real and imaginary parts non-zero, makes  the CPE   a lossy capacitor that sits between the ideal capacitor and the resistor.  
The phase angle is $\phi(\tilde{z})= -\alpha \pi/2 $, constant and independent from frequency, which is why it is called a CPE. The specific case of impedance when $\alpha=1/2$ is known as the Warburg element. 
By integral transform methods, one can carry out the analysis of the CPE behavior in the time domain in terms of voltage, current, and power, which has been done in response to different types and forms of excitations\;\cite{allagui2022time, ieeeted, allagui2024exact, cstRIP}.  Although this compact   formulation of the CPE is invaluable for describing the frequency-dispersed processes in electrochemical systems and devices\;\cite{bisquert2025brief}, that would  otherwise  require a large number of elementary $RC$ circuits, the lack of clear physical understanding of the parameter $\alpha$ is still a limitation to its broader usage.

\begin{figure*}[t]
\begin{center}
\includegraphics[angle=-90,width=0.85\textwidth]{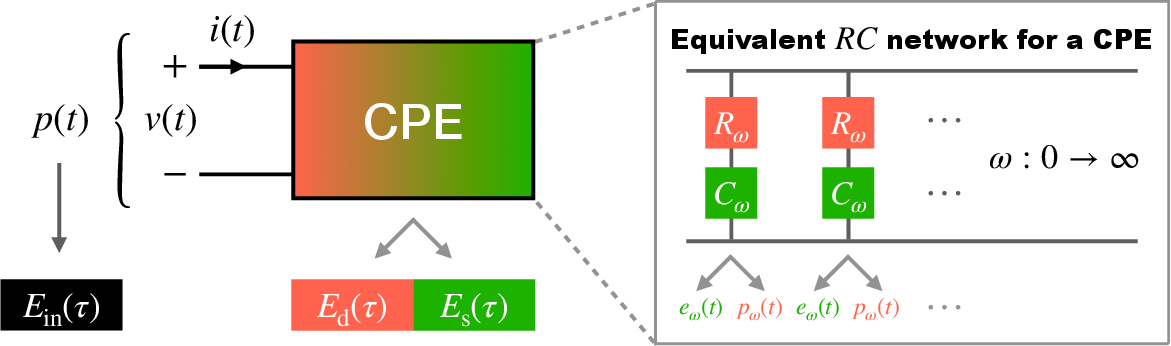}
\caption{Schematic diagram representing the different   energy metrics in a constant phase element (CPE); the source input energy at a time $\tau$ ($E_{{in}}(\tau)$) is equal to the sum of the stored energy in the capacitive part of the CPE 
($E_{s} (\tau)$) and the dissipated energy in its  resistive part ($E_{d}(\tau)$). The inset shows an $RC$ equivalent circuit for a CPE consisting an infinite number of series $R_{\omega} C_{\omega}$ branches connected in parallel, wherein each branch contributes with an elemental or modal stored energy $e_{\omega}(t)$ (in the  capacitor $C_{\omega}$),   and generates an elemental dissipated power $p_{\omega}(t)$ (due to the resistor $R_{\omega}$)}
\label{fig1}
\end{center}
\end{figure*}

The purpose of this work is to propose an energy-based interpretation of  the dispersion coefficient $\alpha$ of a CPE.  This is carried out by splitting the input energy into a contribution that is stored in the capacitive parts of the CPE and another that is dissipated in its resistive parts   using the $RC$ network equivalency.  In particular, we   show that in response to an applied constant voltage, the value $\alpha$ is found to be proportional to the natural logarithm of the input-to-dissipated energy ratio. For the case of a dynamic excitation by a voltage ramp or a voltage $v(t) \propto t^2$, the coefficient $\alpha$ is also found to follow a function of the same energy fraction   (but more complex than the step voltage case), without the involvement of any other parameters such as the pseudocapacitance $c_{\alpha}$ or the magnitude of the voltage scan rate or the duration of the excitation, or others. The CPE coefficient $\alpha$ is therefore not just an empirical fitting parameter in the study of spectral impedance data, but an index that represents the tradeoff between the dissipated and stored energies in response to a given input energy. The results are expected to have broad implications for the analysis of porous electrodes in supercapacitors, batteries, fuel cells and other devices exhibiting the CPE type of response.

\section{Theory}

\subsection{Current-voltage relationship}

The time-domain current-voltage relationship for a CPE is given by the Caputo time-fractional derivative:
\begin{equation}
i(t) = c_{\alpha}\,
{}_0\text{D}_t^{\alpha} v(t)
\label{eq:iCPE}
\end{equation}  
 The Caputo differential operator ${}_0\text{D}_t^{\alpha} f(t)$ of order $\alpha$ ($m-1< \alpha \leqslant m$, $m\in \mathbb{N}$) of a function $f(t)$ is defined starting from its representation in the Laplace   domain by\;\cite{podlubny1998fractional}:
\begin{equation}
\mathcal{L}\left[{}_0\text{D}_t^{\alpha} f(t); s \right]= s^{\alpha} \tilde{f}(s) - \sum\limits_{k=0}^{m-1} s^{\alpha-k-1} f^{(k)}(0^+)
\label{eq:CaputoD}
\end{equation}
where $f^{(k)}(t)$ is the $k^{\text{th}}$ derivative of $f(t)$ with respect to $t$. This leads to the definition  of the Caputo fractional derivative as\;\cite{podlubny1998fractional}:
\begin{equation}
{}_0\text{D}_t^{\alpha} f(t) =  
 \frac{1}{\Gamma(m-\alpha)} \int_0^t (t-\tau)^{m-\alpha-1} f^{(m)}(\tau) d\tau
 \end{equation} 
 for $m-1<\alpha<m$. 
For $ \alpha =m$, we retrieve the standard integer-order derivative ${d^m f(t)}/{dt^m}$. 
When a voltage $v(t)$ is applied onto the CPE, the current $i(t)$ is obtained directly from  Eq.\;\ref{eq:iCPE}, and  
when a current $i(t)$ is applied,  the resulting voltage $v(t)$ can be found from  the inverse Laplace transform  of $ \tilde{z}(s)  \, \tilde{i}(s) $ as:
\begin{equation}
v(t)
=\mathcal{L}^{-1}\left[ \tilde{z}(s)  \,\tilde{i}(s)  ; t \right]
= \frac{1}{c_{\alpha} \Gamma(\alpha)} \int_0^t (t-\tau)^{\alpha-1} i(\tau) d\tau
\end{equation} 
 
From these two quantities, it is straightforward to obtain the impedance of the CPE, $
\tilde{z}(s)
  = {1}/({c_{\alpha} s^{\alpha}})
$, from the ratio of the Laplace transform of the voltage by that of the current. 

\subsection{Input, dissipated and stored energies}

The  instantaneous power input $p(t)$ that goes into a CPE from an external power supply is given by the point-wise product of the voltage by the current in the time domain (see Fig.\;\ref{fig1}) as:
\begin{equation}
p(t)=v(t) i(t)=    c_{\alpha}\, v(t)\, {}_0\text{D}_t^{\alpha} v(t) 
\label{eq:power}
\end{equation}
from which the total energy  up to a time $\tau$ is given by the  integral:
\begin{equation}
E_{\rm in}(\tau) = \int_0^{\tau}  v(t) i(t) dt
\label{eq:Ein}
\end{equation}

In order to account for the portion of this energy that is effectively stored by the capacitive part of the CPE and the portion that is irreversibly dissipated by its resistive part (see Fig.\;\ref{fig1}), the power $p(t)$ in Eq.\;\ref{eq:power} is re-formulated in terms of the infinite state approach as follows\;\cite{hartley2015energy, trigeassou2012state, TRIGEASSOU2013892, yuan2017mechanical, yuan2020mechanical}.
First, the frequency domain operator $s^{\alpha}$ (see Eq.\;\ref{eq:CaputoD}) is rewritten in the form of a Stieltjes transform  as\;\cite{allagui2024procedure, allagui2024generalized}:
\begin{equation}
 s^{\alpha} =   \int_0^{\infty} \mu(\omega) \frac{ s}{s+\omega} d\omega 
 \label{eq:sAlpha}
\end{equation}
where $\mu(\omega)$ is given by:
\begin{equation}
\mu(\omega) = \pi^{-1}{\sin(\alpha \pi)}\, \omega^{\alpha-1},\;\;0<\alpha<1
\label{eq:mu}
\end{equation} 
Then, the fractional derivative $_0\text{D}_t^{\alpha} v(t)$ is obtained from   the inverse Laplace transform of $s^{\alpha} \tilde{v}(s)$ (see Eq.\;\ref{eq:CaputoD}) by using the convolution theorem, such that:
\begin{equation}
_0\text{D}_t^{\alpha} v(t)= \int_0^{\infty} \mu(\omega) \frac{d}{dt} \left[ \int_0^t e^{-\omega(t-\tau)} v(\tau) d\tau \right] d\omega
\label{eq:Dcaputo}
\end{equation}
where we used the Laplace transform pair $(s+\omega)^{-1} \div e^{-\omega t}$. We assumed zero initial conditions, i.e. $v^{(k)}(0^+)=0$ for all $k$ (see Eq.\;\ref{eq:CaputoD}). 
Defining the modal state variable $x_{\omega}(t)$ for the mode $\omega$ as:
\begin{equation}
x_{\omega}(t)=  \int_0^t e^{-\omega(t-\tau)} v({\tau})\, d\tau
\end{equation} 
then, with the use of Leibniz rule for differentiation under the integral sign, we obtain for the time derivative of $x_{\omega}$ in Eq.\;\ref{eq:Dcaputo} the following result:
\begin{equation}
\frac{d x_{\omega}(t)}{dt}=  - \omega x_{\omega}(t) + v(t)
\label{eq:15}
\end{equation}
The CPE power $p(t)$ in Eq.\;\ref{eq:power} can now be expressed as:
\begin{equation}
p(t)= c_{\alpha} 
 \int_0^{\infty} \mu(\omega) \left[ v(t)^2 - \omega x_{\omega}(t) v(t) \right] d\omega
 \label{eq:16}
\end{equation}

This is not just an algebric splitting of the power into two terms. Actually 
each  amount of power for a given mode $\omega$  from Eq.\;\ref{eq:16}, i.e. the integrand:
\begin{equation}
p_{\omega}=c_{\alpha} 
  \mu(\omega) \left[ v(t)^2 - \omega x_{\omega}(t) v(t) \right] 
  \label{eq:pw}
\end{equation}
 can be viewed as the power in an elemental series-connected $R_{\omega}C_{\omega}$ circuit when a voltage $v(t)$ is applied (see insert in Fig.\;\ref{fig1}).  
  For such a circuit, the current is given by the ratio of the voltage across the elemental resistor $R_{\omega}$ by $R_{\omega}$, i.e.:
  \begin{equation}
i_{\omega}(t)=\frac{v(t)-v_{C_{\omega}}(t)}{R_{\omega}}
\end{equation}
 where $v_{C_{\omega}}(t)$ is the voltage across the elemental capacitor $C_{\omega}$. Thus the power in Eq.\;\ref{eq:pw} is also:  
 \begin{equation}
p_{\omega}(t)= \frac{ v(t)^2 - v_{C_{\omega}}(t) v(t) }{R_{\omega} }
\end{equation}
This implies, when comparing with the expression in Eq.\;\ref{eq:pw}, that:
\begin{equation}
R_{\omega}= \left({c_{\alpha} \mu(\omega)}\right)^{-1}, \quad v_{C_{\omega}}(t) = \omega x_{\omega}(t)
\label{eq:19}
\end{equation}
Now from the capacitor's current-voltage  characteristic equation ${i_{\omega}}(t)={C_{\omega}} {dv_{C_{\omega}}(t)}/{dt}$ (with the use of Eqs.\;\ref{eq:19} and\;\ref{eq:15}) we write:
\begin{align}
 \frac{ v(t)  - v_{C_{\omega}}(t)   }{R_{\omega} C_{\omega}} =   \omega \frac{dx_{\omega}(t)}{dt} = \omega (  v(t) - v_{C_{\omega}}(t) )
\end{align}
from which we find the capacitance $C_{\omega}$ in the elemental $R_{\omega}C_{\omega}$ circuit as:
\begin{equation}
C_{\omega}=\left({\omega R_{\omega}}\right)^{-1} = c_{\alpha} \omega^{-1} \mu(\omega) 
\end{equation}
The time constant of each $R_{\omega}C_{\omega}$ branch along the line is $1/\omega$. 
With these parameters, the      stored electrostatic energy in the  capacitor $C_{\omega}$ (denoted $e_{\omega}(t)$) and the dissipated power in the resistor $R_{\omega}$ (denoted $p_{\omega}(t)$) for a given mode $\omega$ or branch $R_{\omega}C_{\omega}$  are:
\begin{align}
&e_{\omega}(t)= \frac{1}{2} C_{\omega} \left[v_{C_{\omega}}(t)\right]^2= \frac{1}{2} c_{\alpha}  \mu(\omega)\,\omega\,  x_{\omega}(t)^2\\
&p_{\omega}(t)= R_{\omega} i_{\omega}(t)^2=  c_{\alpha} \mu(\omega) \left( \frac{d x_{\omega}(t)}{dt}\right)^2
\label{eq:pomega}
\end{align}
respectively (see Fig.\;\ref{fig1}).

After summing over all modes, we obtain the stored energy in the whole CPE at a time $\tau$ as:
\begin{align}
E_{s}(\tau)= \frac{1}{2} c_{\alpha}   \int_0^{\infty}  \mu(\omega) \, \omega \, x_{\omega}(\tau)^2 d\omega    
\label{eq:ECPEsDef}
 \end{align}
and the total (cumulative) dissipated energy in the CPE  as the double integral:
\begin{align}
E_{d}(\tau)= c_{\alpha} \int_0^{\tau} \int_0^{\infty}  \mu(\omega) \left( \frac{d x_{\omega}(t)}{dt}\right)^2 d\omega   dt
\label{eq:ECPEdDef}
\end{align}
The dissipated power, a strictly positive quantity,  
 is of course converted irreversibly into Joule thermal energy, and an increase of the temperature of the system is expected. In principle, it is from this change of temperature that the dissipated energy can be measured experimentally. The stored energy should follow from the energy balance equation:
 \begin{equation}
E_{{  in}}(\tau)=
E_{s} (\tau)+
E_{d}(\tau)
\label{eq:EnergyBalance}
\end{equation} 
 
Note that for the evaluation of $E_{s}$ in Eq.\;\ref{eq:ECPEsDef}, it is useful to use the result:
\begin{equation}
\int_0^{\infty} \omega^{\nu-1} e^{-\omega s} d\omega = \Gamma(\nu) s^{-\nu}
\end{equation}
which gives:
\begin{align}
E_{s}(\tau)&= \frac{c_{\alpha} \sin(\alpha \pi)}{2 \pi} \int_0^{\infty} \omega^{\alpha} \left(\int_0^{\tau} e^{-\omega(\tau-\xi)} v(\xi)\, d\xi \right)^2 d\omega\\
&=\frac{ c_{\alpha} \sin(\alpha \pi)}{2 \pi} \iint_{[0,\tau]^2} v({\tau_1}) v({\tau_2}) \int_0^{\infty} \omega^{\alpha} e^{-\omega(2\tau-\tau_1 - \tau_2)} d\omega d\tau_1 d\tau_2 v\\
&=\frac{\alpha c_{\alpha}}{2 \Gamma(1-\alpha)} \iint_{[0,\tau]^2} v(\tau_1) v(\tau_2) (2 \tau - \tau_1-\tau_2)^{-(\alpha+1)}  d\tau_1 d\tau_2
\label{eq:integralCPE} 
\end{align}
For $E_d(t)$ we use the state variable derivative expression:
\begin{equation}
\frac{d x_{\omega}(t)}{dt}=  \int_0^t e^{-\omega(t-\tau)} \dot{v}({\tau})\, d\tau
\end{equation}
under the assumption of  zero initial condition (obtained by using the differentiation of a convolution integral).
Thus, we have:
\begin{align}
E_{d}(\tau)= \frac{  c_{\alpha}}{  \Gamma(1-\alpha)} \int_0^{\tau} \iint_{[0,t]^2} \dot{v}(\tau_1) \dot{v}(\tau_2) (2 t - \tau_1-\tau_2)^{-\alpha}  d\tau_1 d\tau_2 dt
\label{Eq:Ed3int}
 \end{align}

From  the expressions we have for 
$E_{{  in}}(\tau)$, 
$E_{s} (\tau)$, and 
$E_{d}(\tau)$, 
the ratios:
\begin{align}
&\frac{ E_{s}(\tau) }{E_{{in}}(\tau)} = 1- \frac{   E_{ d} (\tau) }{E_{ {in}} (\tau)} \\
& \frac{ E_{s}(\tau) }{E_{{d}}(\tau)} =  \frac{   E_{in} (\tau) }{E_{ {d}} (\tau)} -1
\end{align}
can be used as  figures of merit for assessing the actual capacitive energy storage efficiency of the CPE relative to its  irreversible dissipated energy.  

In Fig.\;\ref{fig1} we show a schematic diagram summarizing the interplay between the different electrical quantities and circuit elements invoked  in this theoretical analysis, from the macroscopic viewpoint of the CPE to the microscopic elemental $R_{\omega}C_{\omega}$ branches.

\section{Examples}

To facilitate the analytical evaluation of the associated energy quantities $E_{{  in}}(\tau)$, $E_{s} (\tau)$ and $E_{d}(\tau)$ for a CPE, we present below three simple test examples for the applied voltage excitation, that is $v(t) \propto t^{p}$ with $p=0$ (step voltage), $p=1$ (voltage ramp) and $p=2$ (voltage being a quadratic function of time). We examined also the situations of higher orders with  $p=3$ and $p=4$ (results not shown), which led to   the same qualitative conclusions.  

We recall that  the Caputo fractional derivative of a  power law function $f(t)=t^p$ is\;\cite{diethelm2002analysis}:
\begin{equation}
^C_0D_{t}^{\alpha} (t^p) = 
     \frac{\Gamma(1+p)}{\Gamma(1+p-\alpha)} t^{p-\alpha}       
\label{WH}
\end{equation}
if (i) $p \in \mathbb{N}$, $p\geqslant m $ ($m-1<\alpha<m$)  or if (ii)   
$p \notin \mathbb{N}, p > m-1 $.  
Otherwise, if  (iii)  
$  p \in \mathbb{N}$ and $p<m$ we have:
\begin{equation}
    ^C_0D_t^{\alpha} (t^p) =0
\end{equation}
For all other cases, i.e. when (iv) $p<m-1$, $p \neq 0,1,2,\ldots,m-1$, the integral in the expression of the Caputo fractional derivative is improper and divergent\;\cite{tarasov2020exact}. 
 
\subsection{Step voltage}

Assume a constant voltage $v(t)=v_{in} u(t) $ is applied at the time $t=0$ on an initially uncharged CPE. The current is then found from Eq.\;\ref{eq:iCPE},  using Eq.\;\ref{WH} with $p=0$, to be the power law decay:
\begin{equation}
i(t) = \frac{c_{\alpha} v_{in}}{\Gamma(1-\alpha)} t^{-\alpha}
\end{equation}
From Eq.\;\ref{eq:Ein}, the cumulative input energy for the time duration $\tau$ of the excitation is:
\begin{equation}
E_{in}(\tau) = \frac{c_{\alpha} v_{in}^2}{\Gamma(2-\alpha)} \tau^{1-\alpha}
\end{equation}
and  from Eq.\;\ref{eq:integralCPE}, the stored energy at the time $\tau$ is found to be:  
\begin{align}
E_{s}(\tau)
&=
\frac{\alpha c_{\alpha} v_{in}^2}{2 \Gamma(1-\alpha)} \iint_{[0,\tau]^2}  (2 \tau - \tau_1-\tau_2)^{-(\alpha+1)}  d\tau_1 d\tau_2
\\ & =
(1-2^{-\alpha}) \frac{c_{\alpha} v_{in}^2}{\Gamma(2-\alpha)} \tau^{1-\alpha} 
\end{align}
The dissipated energy is deducted from the difference of the two using Eq.\;\ref{eq:EnergyBalance}. 
Then we have the $\alpha$-only-dependent energy ratios:
\begin{align}
&\frac{E_{s}(\tau)}{E_{in}(\tau)}= 1-2^{-\alpha} \label{eq:EsEin}\\
&\frac{E_{d}(\tau)}{E_{in}(\tau)}= 2^{-\alpha} \label{eq:EdEin} \\
&\frac{E_{s}(\tau)}{E_{d}(\tau)}= 2^{\alpha}-1 \label{eq:EsEd} 
\end{align}
from which $\alpha$ can be found to be:
\begin{equation}
\alpha=\frac{\ln\left(\frac{E_{in}(\tau)}{E_{d}(\tau)}\right)}{\ln(2)} = \log_2\left(\frac{E_{in}(\tau)}{E_{d}(\tau)}\right)
\label{eq:alphaCV}
\end{equation}
 It can be readily verified  that at the limit  $\alpha\to 0$, the CPE behaves as a pure resistor given that all of the input energy dissipates into the resistive part of the CPE  (${E_{d}(\tau)}/{E_{in}(\tau)} \to 1$ and ${E_{s}(\tau)}/{E_{in}(\tau)} \to 0$). Conversely,   at the limit $\alpha\to 1$, which corresponds to an ideal capacitor, we have both ${E_{s}(\tau)}/{E_{in}(\tau)}$ and ${E_{d}(\tau)}/{E_{in}(\tau)}$ tending to 1/2, as expected from classical results. For $\alpha=1/2$, we have ${E_{d}(\tau)}/{E_{in}(\tau)} = 1/\sqrt{2} \approx 0.707$ and ${E_{s}(\tau)}/{E_{in}(\tau)}=1- 1/\sqrt{2} \approx 0.293$, which is exactly what is obtained for a  semi-infinite $RC$ transmission line in
response to a voltage step excitation, knowing that the impedance of such a network is in fact of the Warburg type with $\alpha=1/2$.  In Fig\;\ref{fig1} we show plot of of Eqs\;\ref{eq:EsEin}-\ref{eq:EsEd} for $\alpha$ varying between 0.05 and 0.95.

\begin{figure}[t]
\begin{center}
\includegraphics[width=7cm]{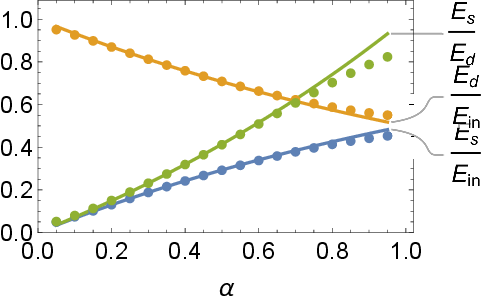}
\caption{Energy metrics in a CPE of different dispersion coefficient $\alpha$ under a constant voltage step excitation; the solid line plots are for Eqs\;\ref{eq:EsEin}-\ref{eq:EsEd} and the dotted plots for the CPE emulator based on the distribution function of relaxation times given by Eq.\;\ref{eq:DFRT}}
\label{fig2}
\end{center}
\end{figure}

This direct connection of the dispersion coefficient $\alpha$ of the CPE to the energy ratio  $E_{d}(\tau)/E_{in}(\tau)$ only (when it is subjected to a step voltage excitation for the moment) is a particularly remarkable result. There is no dependence of the $\alpha$-function  on $v_{in}$, $c_{\alpha}$ or $\tau$ or any other parameters. The value of $\alpha$ for a CPE can therefore be interpreted  as a quantifier for how the input energy splits into dissipative and stored energies within the CPE.

We note that from an experimental standpoint, the dissipated energy in the CPE when subjected to a well-defined electrical excitation  is probably more straightforward  to determine, since it can  be inferred  from the temperature rise of the CPE, and thus directly related to the  Joule heating effect. The stored energy is  then obtained indirectly from the energy balance (Eq.\;\ref{eq:EnergyBalance}), as it cannot be measured directly.

As a way of verifying the results from another perspective, we rewrite  the admittance of a CPE  using Eq.\;\ref{eq:sAlpha} as follows\;\cite{maradesa2024advancing}:
\begin{equation}
\tilde{y}(s)=c_{\alpha} s^{\alpha} = \int_{0}^{\infty} g(\tau) \frac{s \tau}{1+s\tau } d \ln \tau
\label{eq:DFRT}
\end{equation}
where   $g(\tau)$ is given by:
\begin{equation}
g(\tau) = c_{\alpha} \pi^{-1}  \sin(\alpha \pi) \tau^{-\alpha}
\end{equation}
and is known as the distribution function of relaxation times (DFRT)\;\cite{allagui2024procedure}. 
A single parallel $R_{i} C_{i}$ branch in this CPE emulator  network has an elemental admittance as:
\begin{equation}
y_{i}(s)= \frac{1}{R_i} \frac{s \tau_i}{1+ s\tau_i}
\end{equation}
where $\tau_i= R_i C_i$.   The  voltage across each branch capacitor   in response to the step voltage $v_{in} u(t)$ is:
\begin{equation*}
v_{C_i}(t) = v_{in} (1-e^{-t/\tau_i})
\end{equation*}
and the current in each branch is:
\begin{equation}
i_i(t) = \frac{v_{in}}{R_i} e^{-t/\tau_i}
\end{equation}
We take $1/{R_i}= g(\tau_i) \Delta(\ln \tau)$ and therefore $C_i=\tau_i/R_i$.
The total current is found from the sum of all currents passing through each branch. This is then followed by the numerical computation of the total input, stored and dissipated energies for the whole network as done above.  
 The plots of ${E_{s}(\tau)}/{E_{in}(\tau)}$, ${E_{d}(\tau)}/{E_{in}(\tau)}$ and ${E_{s}(\tau)}/{E_{d}(\tau)}$ vs. $\alpha$ (from 0.05 to 0.95 by step of 0.05),  with $\tau=10$  for a network of 1000 $RC$ branches with time constants spaced logarithmically between $10^{-3}$ and $1.6\times 10^{11}$\,s are   shown in dotted markers in Fig.\;\ref{fig2}. They are clearly in excellent agreement with the theoretical values given by Eqs.\;\ref{eq:EsEin}-\ref{eq:EsEd}. The increased difference between the curves for high values of $\alpha$ is due to the decreased accuracy of representing a CPE with Eq.\;\ref{eq:DFRT} (which tends toward the Dirac delta function as $\alpha\to 1$) for the chosen range of relaxation times.
 
\subsection{Voltage ramp}

As a second example, we consider the case of  a voltage ramp $v(t)=\beta t$ where $\beta$ is the voltage scan rate in unit of  V\,s$^{-1}$.   From Eq.\;\ref{WH} with $p=1$ we have the following result for the current along the CPE:
\begin{equation}
i(t)   =  
     \frac{ c_{\alpha} \beta}{\Gamma(2-\alpha)} t^{1-\alpha}       
\end{equation}
Then we obtain the energy quantities $E_{in}(\tau)$, $E_{s}(\tau)$ and $E_{d}(\tau)$ as follows:
\begin{align}
&E_{in}(\tau) = \frac{ c_{\alpha} \beta^2}{ (3-\alpha) \Gamma(2-\alpha)} \tau^{3-\alpha}
\\
&E_{s}(\tau) = (  4 -2^{2-\alpha} -\alpha  ) \frac{ c_{\alpha} \beta^2 }{   \Gamma(4-\alpha)} \tau^{3-\alpha}
\\
&E_{d}(\tau) = (  2^{2-\alpha} -2  ) \frac{ c_{\alpha} \beta^2}{   \Gamma(4-\alpha)} \tau^{3-\alpha}
\end{align}
This last expression of $E_{d}(\tau)$ is  integrated directly using its expression given in Eq.\;\ref{Eq:Ed3int}. We verify that the same  can be deducted from the energy balance equation given by Eq.\;\ref{eq:EnergyBalance}. 

\begin{figure}[b]
\begin{center}
\includegraphics[width=7cm]{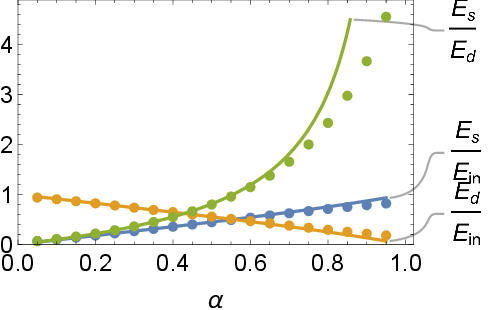}
\caption{Energy metrics in a CPE of different dispersion coefficient $\alpha$ under a  voltage ramp excitation; the solid line plots are for Eqs.\;\ref{eq:EsEinRampV}-\ref{eq:EsEdRampV} and the dotted plots for the CPE emulator based on the distribution function of relaxation times given by Eq.\;\ref{eq:DFRT}}
\label{fig3}
\end{center}
\end{figure}

In Fig\;\ref{fig2} we show plots  of the    energy fractions ${E_{s}(\tau)}/{E_{in}(\tau)}$, ${E_{d}(\tau)}/{E_{in}(\tau)}$ and ${E_{s}(\tau)}/{E_{d}(\tau)}$ given by the expressions:
\begin{align}
&\frac{E_{s}(\tau)}{E_{in}(\tau)}= \frac{4-2^{2-\alpha}-\alpha}{2-\alpha} \label{eq:EsEinRampV}
\\
&\frac{E_{d}(\tau)}{E_{in}(\tau)}=\frac{2^{2-\alpha}-2}{2-\alpha} \label{eq:EdEinRampV}
 \\
&\frac{E_{s}(\tau)}{E_{d}(\tau)}= \frac{4-2^{2-\alpha}-\alpha}{2^{2-\alpha}-2} 
\label{eq:EsEdRampV}
\end{align} 
for $\alpha$ varying  between 0.05 and 0.95. These ratios are still dependent on the coefficient  $\alpha$ only, but the functional forms are more complex than those derived for the constant voltage case. The dotted curves represent the energy ratios obtained using the analysis derived from Eq.\;\ref{eq:DFRT} with the appropriate expressions of branch capacitor voltage and branch current in response to a ramp excitation. They also reproduce the theoretical trends of Eqs.\;\ref{eq:EsEinRampV}-\ref{eq:EsEdRampV} with very reasonable fidelity. 
 We note the limiting cases: 
${E_{d}(\tau)}/{E_{in}(\tau)} \to 1$ as $\alpha \to 0$ (all input energy is dissipated), 
${E_{d}(\tau)}/{E_{in}(\tau)} \to 0$ as $\alpha \to 1$ (all input energy is stored), and ${E_{d}(\tau)}= {E_{s}(\tau)}$ takes place at $\alpha=0.555$ (see Fig\;\ref{fig2}). 
 
The solution of the transcendental equation given by Eq.\;\ref{eq:EdEinRampV} for $\alpha$, that we rewrite as:
\begin{equation}
\frac{2^u-2}{u}=k
\end{equation}
where $u=2-\alpha$, $2^u=e^{u \ln(2)}$ and $k={E_{d}(\tau)}/{E_{in}(\tau)}$ is found to be:
\begin{equation}
u= 2-\alpha=-\frac{2}{k} - \frac{W[-4^{-1/k} k^{-1}\ln(2)]}{\ln(2)}
\end{equation}
Here $W(z)$ is the Lambert function. In this case of voltage ramp excitation, the expression of $\alpha$ is still a pure function of the energy   ratio ${E_{d}(\tau)}/{E_{in}(\tau)}$, but different from the simple relation  found for a constant voltage excitation (Eq.\;\ref{eq:alphaCV}). The reason  is because the input voltage is a dynamic function of time, which implies that $\alpha$  takes into account both the fractional dynamics of the CPE and the fact that  the input energy  keeps changing  via the linearly increasing   voltage.

\subsection{Voltage $v(t)=\lambda t^2$}

\begin{figure}[b]
\begin{center}
\includegraphics[width=7cm]{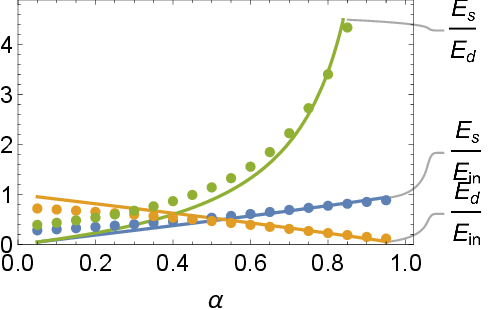}
\caption{Energy metrics in a CPE of different dispersion coefficient $\alpha$ under a voltage  excitation of the form $v(t)=\lambda t^2$; the solid line plots are for Eqs.\;\ref{eq:EsEinV2}-\ref{eq:EsEdV2} and the dotted plots for the CPE emulator based on the distribution function of relaxation times given by Eq.\;\ref{eq:DFRT}}
\label{fig4}
\end{center}
\end{figure}

As a third example, we examine the case of quadratic voltage input of the form $v(t)=\lambda t^2$ where $\lambda$ is a pure constant. While this  type of signal is unusual, it is employed here to extend the validation of our analysis to a nonlinear excitation, allowing us to demonstrate that the conclusions remain consistent beyond the simple voltage step and ramp inputs from above. We find in this case that the energy ratios are also given by $\alpha$-only-dependent functions as follows:
\begin{align}
&\frac{E_{s}(\tau)}{E_{in}(\tau)}= \frac{\alpha^2-11\alpha +2^5-2^{5-\alpha}}{(\alpha-3)(\alpha-4)}
\label{eq:EsEinV2}
\\
&\frac{E_{d}(\tau)}{E_{in}(\tau)}=\frac{4(\alpha-5+2^{3-\alpha})}{(\alpha-3)(\alpha-4)}  
 \\
&\frac{E_{s}(\tau)}{E_{d}(\tau)}= \frac{\alpha^2-11\alpha +2^5-2^{5-\alpha}}{4(\alpha-5+2^{3-\alpha})}
\label{eq:EsEdV2} 
\end{align} 
which are all plotted in Fig.\;\ref{fig4} for the same range of $\alpha$. 
Again, the dotted curves represent the energy ratios obtained using Eq.\;\ref{eq:DFRT}, with the appropriate expressions of branch capacitor voltage and branch current in response to this quadratic voltage excitation. We used in this case the minimum and maximum of the time constants to be $10^{-5}$ and $100$\,s, respectively. 
The trends of the curves are overall similar to those obtained with the preceding case of voltage ramp. We note the limiting scenarios: (i) for $\alpha=1$, ${E_{s}(\tau)}/{E_{in}(\tau)}=1$ and ${E_{d}(\tau)}/{E_{in}(\tau)}=0$, whereas (ii) for 
$\alpha=0$, ${E_{s}(\tau)}/{E_{in}(\tau)}=0$ and ${E_{d}(\tau)}/{E_{in}(\tau)}=1$. (iii) The values of $\alpha$ corresponding to equal amounts of  stored and  dissipated energies is found numerically to be $\alpha=0.529$.

\section{Conclusion}
\label{section:conclusion}

From the analysis of three test examples of voltage excitations applied on a CPE (static dc voltage, and dynamic voltage of ramp and quadratic forms), we have shown that the energy fractions ${E_{s}(\tau)}/{E_{in}(\tau)}$, ${E_{d}(\tau)}/{E_{in}(\tau)}$ and ${E_{s}(\tau)}/{E_{d}(\tau)}$ after a time duration $\tau$ are all  dependent solely  on the dispersion coefficient $\alpha$, and nothing else (i.e. independent of amplitude of input signal, duration of excitation, materials-specific parameters, etc.). The functional form varies with the type of the input excitation.  Therefore, the coefficient $\alpha$ is not just a fitting parameter for impedance spectroscopy data, but can now be viewed as an energetic quantifier  (or divider, or index) that indicates how     the  input energy supplied to the CPE is   partitioned between storage and dissipation. In other words, it an indicator of the energy storage efficiency of the CPE, bridging the impedance spectroscopy  of capacitive systems with their thermodynamic energy balance. 

From the experimental side, the estimation of the dissipated energy in the CPE for a given electrical excitation (decoupled from the stored energy) can be obtained directly from the rise in temperature of the CPE   and the associated Joule heating using standard calorimetric techniques. The  stored energy is then obtained by deduction from the energy balance equation, as it may be difficult to measure directly. 

Finally, it is worth mentioning that beyond their  overall conceptual value, the results of this study in decoding the energy partitioning  in a CPE represent  a clear framework that can be readily  applied in supercapacitor analysis, battery health monitoring, and the characterization of other (bio)electrochemical systems exhibiting the CPE behavior.

\section*{Data Availability}


The data that supports the findings of this study are available within the article.


%

\end{document}